\documentclass[aps,prd,showpacs,12pt]{revtex4-1}
\usepackage{graphicx}
\usepackage{amsmath}
\input{epsf}
\begin{document}
\epsfverbosetrue
\def\la{{\langle}}
\def\ra{{\rangle}}
\def\vep{{\varepsilon}}
\newcommand{\beq}{\begin{equation}}
\newcommand{\eeq}{\end{equation}}
\newcommand{\beqa}{\begin{eqnarray}}
\newcommand{\eeqa}{\end{eqnarray}}
\newcommand{\q}{\quad}
\newcommand{\A}{|\Omega'|}
\newcommand{\AC}{{\it AC }}
\newcommand{\n}{\\ \nonumber}
\newcommand{\om}{\omega}
\newcommand{\e}{\epsilon}
\newcommand{\Om}{\Omega}
\newcommand{\os}[1]{#1_{\hbox{\scriptsize {osc}}}}
\newcommand{\cn}[1]{#1_{\hbox{\scriptsize{con}}}}
\newcommand{\sy}[1]{#1_{\hbox{\scriptsize{sys}}}}
\title{Non-Markovian effects in the growth of a polymer chain}
\author {D. Sokolovski$^{a,b}$}
\author {S. Rusconi$^{c}$}
\author {E. Akhmatskaya$^{c,b}$}
\author {J. M. Asua$^f$}
\affiliation{$^a$ Departmento de Qu\'imica-F\'isica, Universidad del Pa\' is Vasco, UPV/EHU, Leioa, Spain}
\affiliation{$^b$ IKERBASQUE, Basque Foundation for Science, E-48011 Bilbao, Spain}
\affiliation {$^c$ Basque Center for Applied Mathematics (BCAM),\\ Alameda de Mazarredo, 14 48009 Bilbao, Bizkaia, Spain}
\affiliation{$^f$ POLYMAT, University of the Basque Country UPV/EHU, Joxe Mari Korta zentroa, Tolosa etorbidea 72, Donostia, San Sebasti‡n 20018, Spain}
\date{\today}
\begin{abstract}
\noindent
{\it ABSTRACT}: 
\newline
Using a simple exactly solvable model, we show that event-dependent time delays may lead to significant non-Poisson effects in the statistics of polymer chain growth. The results are confirmed by stochastic simulation of various growth scenarios.
\end{abstract}
\date{\today}
\maketitle
\section{Introduction}
We consider the growth of a linear polymer chain which proceeds by adding new monomers. 
One may expect the growth to be a Markovian process, with the chain length at a given time being a random quantity distributed according to the Poisson law.  This law will, however, change if acquisition of the next monomer is delayed,
 as it has been
proposed to explain the decrease of the polymerization rate in reversible addition fragmentation
chain transfer (RAFT) controlled radical polymerization \cite{TX1}.
 One can also assume that the growth of the chain can change direction, by starting a branch through intramolecular transfer to polymer as occurs in the free radical polymerization of
acrylates \cite{TX2} and ethylene \cite{TX3},
 but only after its linear segment has reached a certain length. In this case, the appearance of the next branch can be seen as a delayed process, whose delay is approximately the time in which a segment reaches the length required for the branching reaction to be able to take place. In this paper we consider a simple exactly solvable model of a delayed growth. 

The rest of the paper is organised as follows. In Sect. II we define the probabilities for different growth scenarios. 
In Sect.III we briefly review growth without a delay, leading to Markovian master equations for the probabilities of interest. In Sect. IV we consider linear growth with a fixed 'downtime' introduced after each attachment of a monomer. We show that such a delay leads to non-Poisson distribution of the polymer length, and a set of time delayed differential equations for the relevant probabilities. In Sect. V we study the short- and the long-time limits of the mean length of the grown polymer. In Sect. VI we consider a branching process, whose delay is determined by the polymer's growth rate. In Sect, VII we confirm and extend our results by employing a numerical stochastic algorithm similar to the one pioneered by Gillespie \cite{GILL}. Section VIII contains our conclusions.

\section{Linear growth}
We start with the usual mathematical pre-requisit.
Consider the growth of a linear polymer which proceeds by attaching monomers to, say, its right end at discreet times $t_j=jdt$. 
\newline
We begin with a single monomer.
At each $t_j$ an extra monomer is added with the probability $p_j$, 
or else nothing happens with the probability $1-p_j$. Thus, the probability to add a monomer after $J-1$ unsuccessful attempts is $f(t_J)=p_J\prod_{j=1}^{J-1}(1-p_j)$, while the probability for not adding a monomer up to and including $t_J$, is $g(t_J)=\prod_{j=1}^{J}(1-p_j)$. In general, we may start the process at some $t_i$ and let the probabilities depend on both $t_j$ and $t_i$, $p=p_{j,i}$.
In the continuum limit we should send $dt \to 0$ and introduce the growth rate $c(t_j, t_i)dt\equiv p_{j,i}$, assuming $c$ to be a slowly varying function. The probability to add nothing for $t'\le t'' \le t$ is then given by 
 \begin{eqnarray}\label{a1}
g(t,t')=\exp[-\int_{t'}^tc(t'',t')dt''],
\end{eqnarray}
while for the probability density function (PDF) to start at $t'$, and add the first  monomer in the interval $[t, t+dt]$, 
 we have 
 \begin{eqnarray}\label{a2}
f(t,t')=c(t,t')\exp[-\int_{t'}^tc(t'',t')dt''] = -\partial_t g(t,t').
\end{eqnarray}
By a given $t$, the monomer is either attached or not, so the two corresponding probabilities add to one, 
 \begin{eqnarray}\label{a3}
\int_{t'}^t f(t'',t')dt''+g(t,t')=1.
\end{eqnarray}
\newline
With many monomers able to join the polymer chain between $t'$ and $t$, we are looking for the probability $P(n,t,t')$ to have $n$ new additions by the time $t$. This is just the probability for 
adding monomers at $t'\le t_1\le t_2\le...\le t_n\le t$, multiplied by the probability that no more monomers are added between $t_n$ and $t$, and summed over all $t_i$, $i=1,2,..n$, 
 \begin{eqnarray}\label{a4}
P(n,t,t')=\int_{t'}^t dt_n...\int_{t'}^{t_2}dt_1 g(t,t_n)f(t_n,t_{n-1})...f(t_1,t')
, \q n\ge 1\n
P(0,t,t')=g(t,t').
\end{eqnarray}
One can check that Eq.(\ref{a3}) ensures the correct normalisation of the probabilities $P(n,t,t')$, 
$\sum_{n=0}^\infty P(n,t,t')=1$.
\newline
Another useful quantity is the probability density $W(n,t,t')$ for $n$ monomers to be attached in the interval $[t',t]$, with the last of them added in $[t,t+dt]$,  
 \begin{eqnarray}\label{a5}
W(n,t,t')=\int_{t'}^t dt_{n-1}...\int_{t'}^{t_2}dt_1 f(t,t_{n-1})...f(t_1,t')
, \q n\ge 1,\n
\end{eqnarray}
in terms of which $P(n,t,t')$ is expressed as
 \begin{eqnarray}\label{a6}
P(n,t,t')
=\int_{t'}^t dt''g(t,t'')W(n,t'',t'), \q n\ge 1.
\end{eqnarray}
The quantities $W(n,t,t')$ have the advantage that they satisfy the simple evolution equations, 
 \begin{eqnarray}\label{a7}
\partial_t W(n,t,t')=f(t,t)W(n-1,t,t')+\int_{t'}^t dt''\partial_t f(t,t'')W(n-1,t'',t').
\end{eqnarray}
Their use will be described below.
\newline
At least three cases need to be distinguished.
\newline
{\it A.} The growth rate depends only on the current time, and not on the previous history of the chain,
 \begin{eqnarray}\label{a8}
c(t,t')=c(t), \q f(t,t')=c(t)\exp[-\int_{t'}^tc(t'')dt'']. 
\end{eqnarray}
For example, an increase in the temperature may make the attachment of monomers more probable at later times.
This is the {\it Markovian} case we will briefly review in the next Section.
\newline
{\it B.} The growth rate depends only on the chain's past, and is not manipulated externally.
 \begin{eqnarray}\label{a8}
c(t,t')=c(t-t'), \q f(t,t')=f(t-t')=c(t-t')\exp[-\int_{0}^{t-t'}c(t'')dt'']. 
\end{eqnarray}
Here one may think that after each time a monomer is added, some additional time is needed before the next  monomer can be attached \cite{TX1}. This the {\it non-Markovian} case is the main subject of this paper.
\newline
{\it C.} Finally, the growth rate, which depends on the polymer's history in the sense outlined above, may also be manipulated externally. In this case $c$ is a function of both $t$ and $t'$, and the process is also non-Markovian.
\section{Markovian growth}
Consider the case when there is an unlimited supply of monomers to be added to the chain, and the probability to add one at a given time is modified externally, e.g., 
by varying the temperature at which the process takes place.
The growth begins at some $t'$, and we are interested in the length of a polymer at a time $t$. The probability for adding a monomer in $[t,t+dt]$ is $c(t)dt$, and the function $g$ and the PDF $f$ in Eqs. (\ref{a1}) and (\ref{a2}) are of the form
 \begin{eqnarray}\label{b1}
g(t,t')=\exp[-\int_{t'}^tc(t'')dt''],\n
f(t,t')=c(t)\exp[-\int_{t'}^tc(t'')dt''].
\end{eqnarray} 
Inserting (\ref{b1}) into Eq.(\ref{a4}), and recalling  that $\int_{t'}^t dt_n...\int_{t'}^{t_2}dt_1c(t_n)..c(t_1)=[\int_{t'}^tc(t'')dt'']^n/n!$, we recover a Poisson distribution \cite{Pois} (the subscript $M$ stand for 'Markovian'), 
\begin{eqnarray}\label{b2}
P_M(n,t,t')=\frac{I(t)^n}{n!}\exp(-I(t)), \q n\ge1\n
P_M(0,t,t')=\exp(-I(t)),
\end{eqnarray}
where
\begin{eqnarray}\label{b3}
I(t)=\int_{t'}^t c(t'')dt''.
\end{eqnarray}
From Eqs. (\ref{b2})-(\ref{b3}) it follows that the mean length of the chain, 
\begin{eqnarray}\label{b4}
\la n(t)\ra=\bar{c}(t,t')(t-t'), 
\end{eqnarray}
where $\bar{c}(t,t')=(t-t')^{-1}\int_{t'}^t c(t'')dt''$ is the average of the growth rate $c$ over the growth period. If the external conditions remain unchanged, $c(t)=const$, the growth is linear with time,  
$\la n(t)\ra=c(t-t')$.
In the special case (\ref{b1}), differentiating Eq.(\ref{a6}) [or, directly, Eq.(\ref{b2})] yields a closed master equation for the probabilities $P_M(n,t,t')$
\begin{eqnarray}\label{b5}
\partial_tP_M(n,t,t')=c(t)[P_M(n-1,t,t')-P_M(n,t,t')],\q n\ge1,\n
 \partial_tP_M(0,t,t')=-c(t)P_M(0,t,t'),
\end{eqnarray}
to be solved with the initial condition 
\begin{eqnarray}\label{b6}
P(n,t',t')=\delta_{n0}, 
\end{eqnarray}
where $\delta_{nm}$ is the Kronecker delta.
Equations (\ref{b5}) are obviously Markovian, as the rate at which a $P_M(n,t,t')$ changes depends only on the current state of the system, $\{P(n,t,t') \}$, $n=0,1,2...$.
\section{Non-Markovian  growth with delays}
Suppose next that, as in the previous Section, there is an unlimited supply of monomers, and the external conditions remain unchanged. But each added monomer, except the first, now needs a time $\tau$ to properly settle into the chain structure, only after which the chain is ready to attach again, with the same constant growth rate $c$.
The process is now explicitly non-Markovian: to check whether a monomer can be added, one needs to know the history of the chain. Accordingly, the probability $c(t)$ depends not on the time elapsed since $t'$, but on the time elapsed since the last monomer was added. In Eq.(\ref{a4}) we, therefore have
$c(t,t')=c(t-t')$.
 Explicitly, we obtain ($NM$ stands for 'Non-Markovian')
\begin{equation}\label{c1}
c(t-t')=
\begin{cases}
0, & 0\le t-t' < \tau \\
c, & t-t' \ge \tau
\end{cases}
\end{equation}
and
\begin{equation}\label{c2}
g_{NM}(t,t')=g_{NM}(t-t')=
\begin{cases}
1, & 0\le t-t' < \tau \\
\exp[-c(t-t'-\tau)], & t-t' \ge \tau.
\end{cases}
\end{equation}
From (\ref{a2}) we also have
 \begin{eqnarray}\label{c3}
f_{NM}(t,t')=f_{NM}(t-t')=c\theta(t-t'-\tau)\exp[-c(t-t'-\tau)],
\end{eqnarray} 
where $\theta(z)=1$ for $z\ge 0$ and $0$ otherwise.
\newline
It is easy to see that the model described by Eqs.(\ref{c1})-(\ref{c3}) has a simple exact solution.
Indeed, returning to Eq.(\ref{a5}) and putting $t'=0$, we note that the probability $W(n,t,t'=0)\equiv W(n,t)$ is the same as for growth with a constant $c$, but for a shorter time. The effective time of growth, $t_{eff}$, is, therefore, the elapsed time $t$ minus the total time the growth was shut down due to adding $n-1$ monomers, i.e., $t_{eff}=t-(n-1)\tau$. Should $(n-1)\tau$ exceed $t$, the process is not possible, and the corresponding probability is zero. From (\ref{a5}) and (\ref{c3}) we easily find  
 \begin{eqnarray}\label{c4}
W_{NM}(n,t|\tau)=c\frac{[I_n(t,\tau)]^{n-1}}{(n-1)!}\exp[-I_n(t,\tau)],\q n\ge 1, \q (n-1)\tau<t ,
\n
I_n(t,\tau)=c[t-(n-1)\tau].\n
\end{eqnarray}
The physical probabilities $P(n,t,t'=0)\equiv P(n,t)$ are no longer given by a Poisson distribution, but can be obtained as quadratures using Eqs.(\ref{a6}), (\ref{c2}) and (\ref{c4})
 \begin{eqnarray}\label{c5}
P_{NM}(n,t|\tau)=\int_0^{t-\tau}\exp[-c(t-t'-\tau)]W_{NM}(n,t'|\tau)dt'+ \int_{t-\tau}^t W_{NM}(n,t'|\tau)dt', \q n\ge 1, \q\q\n
P_{NM}(0,t|\tau)=\exp(-ct).
\end{eqnarray}
It is instructive to look at the evolution equations (EE) satisfied by the probabilities. There are no simple EE, similar to Eqs.(\ref{b5}), for the $P_{NM}$'s in Eqs.(\ref{c5}). There are, however, EE  (\ref{a7}) which, since $\partial_tf(t)=c\delta(t-\tau)-cf(t)$, read
 \begin{eqnarray}\label{c6}\nonumber
\partial_tW_{NM}(n,t|\tau)=c[W_{NM}(n-1,t-\tau|\tau) - W_{NM}(n,t|\tau)],\q n\ge2,
 \q 0<(n-1)\tau < t.\\
W_{NM}(1,t)=c\exp(-ct).\q\q\q\q\q
\end{eqnarray}
Unlike Eqs.(\ref{b5}) in the Markovian case, Eqs.(\ref{c6}) depend on the state of the system in the past through $W_{NM}(n-1,t-\tau|\tau)$. Since for any suitable function $F(t)$, $F(t-\tau)=\exp(-\tau\partial_t) F(t)=\sum_{n=0}^\infty(-1)^n \frac{\tau}{n!}\partial_t^nF(t)$, Eqs.(\ref{c6}) are, effectively, of infinite order in the time derivative $\partial_t$, and their properties may differ significantly from those of (\ref{b5}), as will be illustrated in the next Section.
\section{mean chain length for a delayed growth}
One quantity of practical interest is the mean length of the chain grown in the presence of a delay,
 \begin{eqnarray}\label{d1}
\la n(t,\tau)\ra_{NM}=\sum_{n=1}^\infty n P_{NM}(n,t|\tau),
\end{eqnarray} 
shown in Fig.1 for various values of the parameters $c\tau$. Figure 2 shows the standard deviation of the length for $c\tau=10$.
\begin{figure}
	\centering
		\includegraphics[width=12cm,height=12cm]{{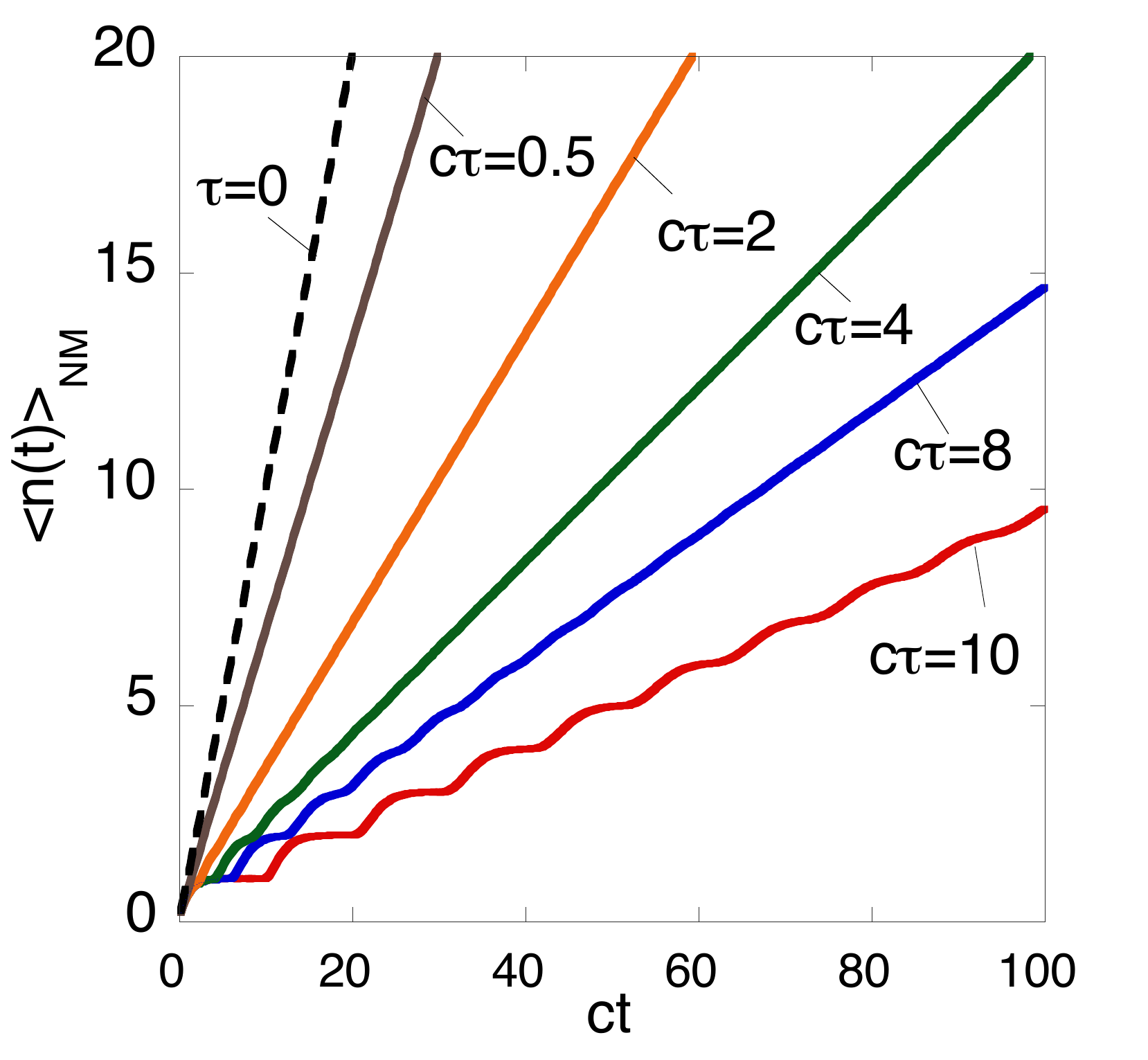}}
\caption{(Color online) The mean length of the polymer chain
  vs. time for
different values of the delay $\tau$.}
\label{fig:4}
\end{figure}
\begin{figure}
	\centering
		\includegraphics[width=12cm,height=12cm]{{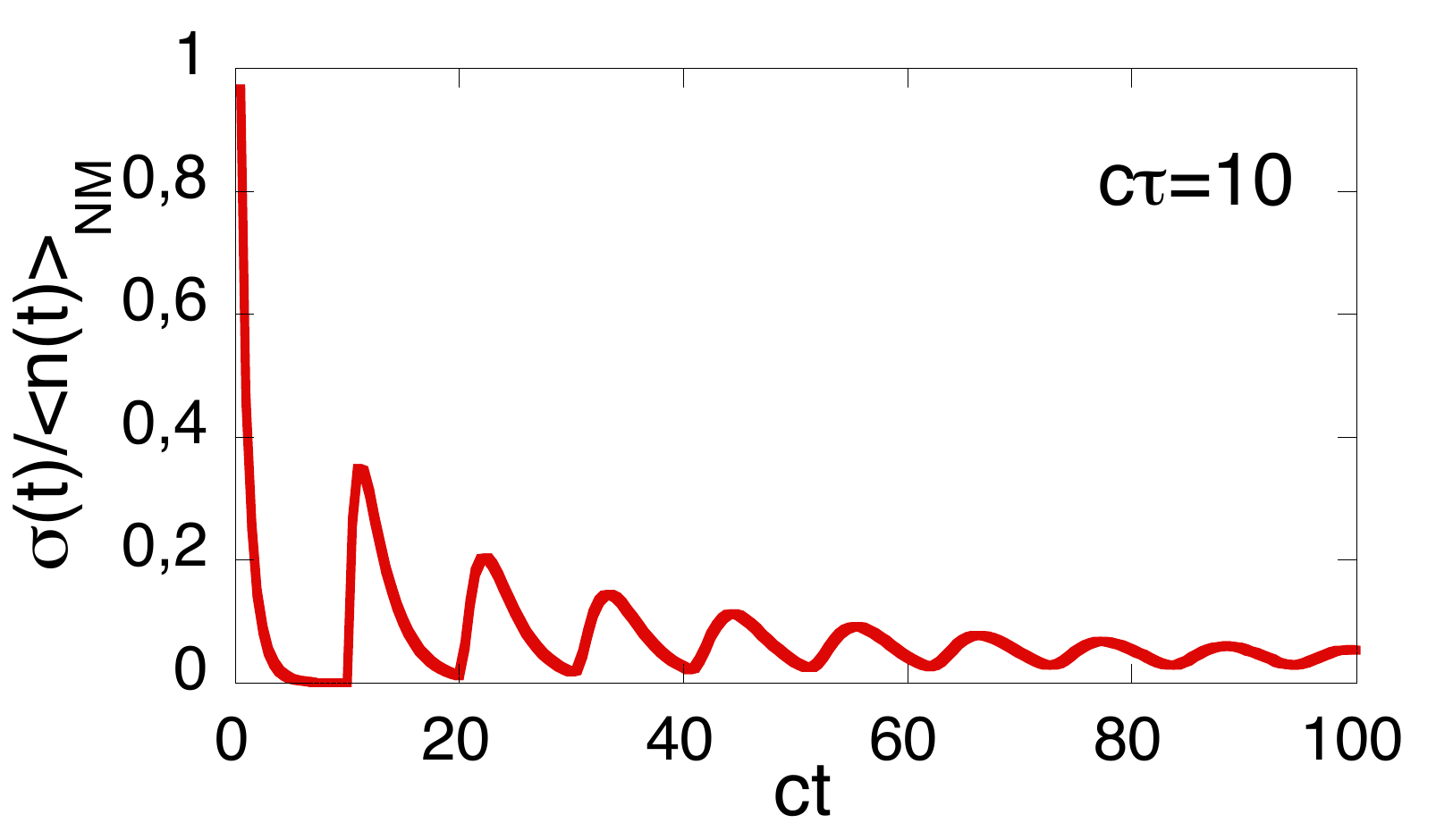}}
\caption{(Color online) The standard deviation of the chain's length, $\sigma \equiv \sqrt{\la n^2\ra_{NM}-\la n\ra_{NM}^2 }$ 
  vs. time for
 $c\tau=10$.}
\label{fig:4}
\end{figure}
As $c\tau \to 0$ we recover the Poisson distribution (\ref{b2}), and for $ct<<1$, $\la n(t,\tau)\ra$ grows linearly at the rate close to $c$. 
In the opposite limit, $ct>>1$, the behaviour is more interesting with the curves showing a steplike variation at short times before settling into a linear behaviour later. This has a simple physical explanation. The time it takes the chain to add one monomer at a constant growth rate $c$, $t_{add}$, is approximately $1/c$. If $\tau/t_{eff} = c\tau >> 1$, a monomer is added quickly, but then the system has to wait long until another one can be attached. Thus, for $t_{add}<t<\tau$, it behaves as if the delay were infinite, i.e. as if the only two possible outcomes were one or none monomers added, with the probabilities $P(one)=1-exp(-ct)$ and $P(none)=exp(-ct)$,  respectively. The mean length 
\begin{eqnarray}\label{d2}
\la n(t,\tau)\ra_{NM}\approx 1-exp(-ct), \q t_{add} <t < \tau
\end{eqnarray}
reaches the value of $1$, and remains unity until $t\approx \tau$, when the system 'recalls' that the delay is not infinite after all. A second monomer is added quickly, and $\la n(t,\tau)\ra_{NM}$ remains flat and close to $2$ until $t\approx 2\tau$, and so on. This behaviour can be expected from the way we have constructed our model. Equivalently, it can be seen as an illustration of the ability of time delayed differential equations to produce rapid variations in their solutions after they seem to have reached an asymptotic limit \cite{DDE}.
\newline
As $t/\tau \to \infty$ the steps are smoothed out, and the mean length of the polymer grows linearly with time ($K$ is a constant), 
 \begin{eqnarray}\label{d3}
\la n(t,\tau)\ra_{NM} \approx \tilde{c}t+K, \q t\to \infty
\end{eqnarray}
at a constant rate $\tilde{c}$, $1/\tau \le \tilde{c} \le c$. The value of $\tilde{c}$ is found by recalling that in our model adding $\la n(t)\ra$ monomers is accompanied by switching off  the growth for a duration of approximately $[\la n(t)\ra-1]\tau - \tau/2$. (The last addition may occur close to $t$, so its delay is, on average, shorter.) Thus, the growth is similar to the growth without delay over a time 
$\la n(t)\ra\tau -\tau/2$ at the rate $c$. Equating $\la n(t)\ra$ to $c[t-\la n(t)\ra\tau +\tau/2]$ yields
 \begin{eqnarray}\label{d4}
\tilde{c}=\frac{c}{1+c\tau},\q K=\frac{c\tau}{2(1+c\tau)}.
\end{eqnarray}
The two limiting cases, (\ref{d2}) and (\ref{d3}), are illustrated in Fig.3. 
\begin{figure}
	\centering
		\includegraphics[width=12cm,height=12cm]{{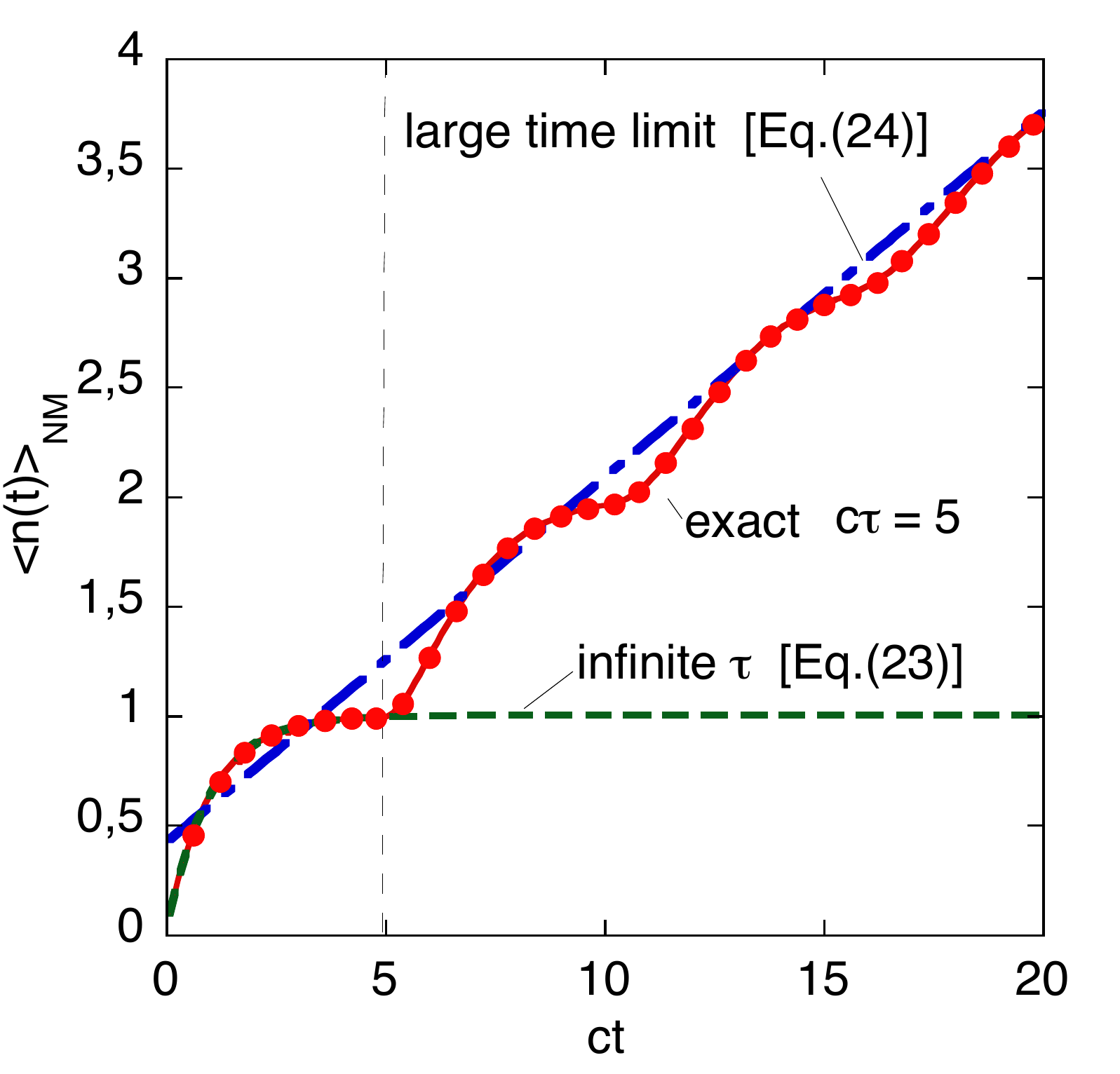}}
\caption{ The mean length of the polymer chain vs. time for $c\tau=5$ (solid). Also shown are the short time limit (\ref{d2}) (dashed) and the long time limit (\ref{d3}) (dot-dashed). 
The filled dots are the results of stochastic simulation with the PDF (\ref{c3}) described in Sect. VII.}
\label{fig:4}
\end{figure}
\section{Linear growth with branching. Growth-induced delay}
Next we consider a different model, in order to illustrate how a delay may arise in practise. The polymer chain grows at a constant rate $c_{add}$, without any delay, its full length at a time $t$ being $c_{add}t$. 
\begin{figure}
	\centering
		\includegraphics[width=9cm,height=6cm]{{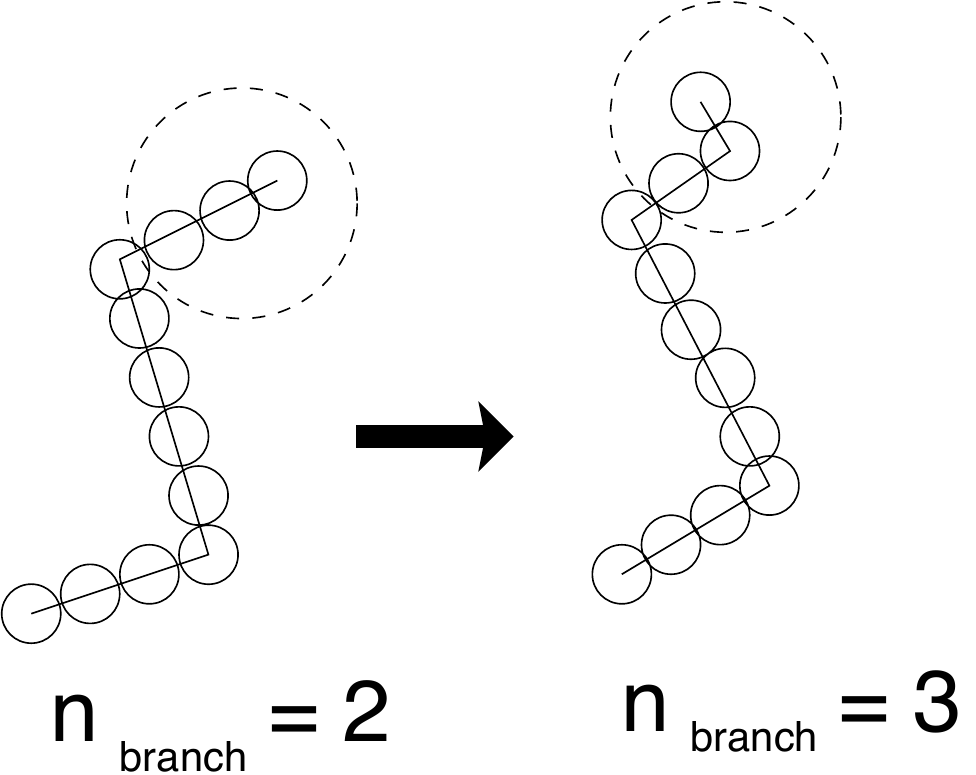}}
\caption{(Color online) A polymer chain grows by adding monomers to its right hand side end at a constant rate $c_{add}$.
The chain also forms branches, at a constant rate $c_{branch}$. After a branching event the chain needs to acquire at least two more monomers until next branching can occur. }
\label{fig:4}
\end{figure}
It can, however, form branches \cite{TX2}, \cite{TX3}
as illustrated in Fig.4.  The branching occurs with a rate $c_{branch}$, but can only happen after a linear segment contains at least three monomers \cite{TX4}. In other words, a branching event should be preceded by at least $n_0=2$ attachments of monomers \cite{FOOT}.
It is the number of brancings, $n_{branch}$, we are interested in. To make the problem tractable, we assume that the growth is deterministic, i.e., that exactly one monomer is added to the chain at $t_{add}=1/c_{add}$. This makes branching  a delayed reaction of the type considered in the Sections IV-V,  and the delay time is now given by
 \begin{eqnarray}\label{e1}
\tau= n_0/c_{add}=n_0 t_{add}.
\end{eqnarray}
The model, which should work well for a large $n_0$, since a branching event can occur just before or just after attachment of an extra monomer, so that the actual delay may lie between $(n_0-1)t_{add}$ and $n_0t_{add}$. We will address this issue shortly.
For now, the mean number of branchings in the chain, $\la n_{branch}(t)\ra$  is given by a formula similar to Eq.(\ref{d2}), but shifted by $\tau$, since already the first branching may occur only after at least $n_0$ growth events, 
 \begin{eqnarray}\label{e2}
\la n_{branch}(t)\ra=\la n(t-n_0t_{add},n_0t_{add})\ra_{NM}.
\end{eqnarray}
The mean number of branchings in the chain, and the mean length of its linear segment, 
 \begin{eqnarray}\label{e3}
\la N(t)\ra \equiv \frac{c_{add}t}{\la n_{branch}(t)\ra+1},
\end{eqnarray}
depend on the ratio
 \begin{eqnarray}\label{e4}
\gamma \equiv n_0 \frac{c_{branch}}{c_{add}}.
\end{eqnarray} 
\newline
For $\gamma >> 1$ and $n_0/c_{add}+1/c_{branch}<t<2n_0/c_{add} $ there is exactly one branching, the second one appearing approximately after $t=2n_0/c_{add}+1/c_{branch}$.
\newline 
For $\gamma >> 1$ and $t\to \infty $ the number of branches grows linearly at the rate obtained from Eqs.(\ref{d4}) and (\ref{e1}):
 \begin{eqnarray}\label{e5}
\tilde{c}_{branch}= \frac{c_{branch}c_{add}}{c_{add}+n_0c_{branch}},
\end{eqnarray}
and 
 \begin{eqnarray}\label{e6}
\la N(t)\ra=c_{add}/\tilde{c}_{branch}=n_0+\frac{c_{add}}{c_{branch}}.
\end{eqnarray}
For $c_{branch} >> c_{add}$ a branching event occurs just after a linear segment has grown to contain $n_0$ monomers, so that $\tilde{c}_{branch}=c_{add}/n_0$  and $\la N(t)\ra=n_0$. 
In the opposite limit $\gamma << 1$ growth outpaces branching.
After an initial delay of $n_0/c_{add}$, branching proceeds at a rate $\approx c_{branch}$, and the mean length of the linear segment, $\la N(t)\ra$, is approximately  
$c_{add}/c_{branch}$.

Figure 5 shows the dependence of $\la N(t)\ra$ on $t$, for $n_0=3$ and different values of $\gamma$ in Eq.(\ref{e4}). Figure 6 shows a related quantity, the ratio of the mean number of the branchings to the mean total length of the chain, $\la n_{branch}(t)\ra/c_{add}t$.
\begin{figure}
	\centering
		\includegraphics[width=12cm,height=12cm]{{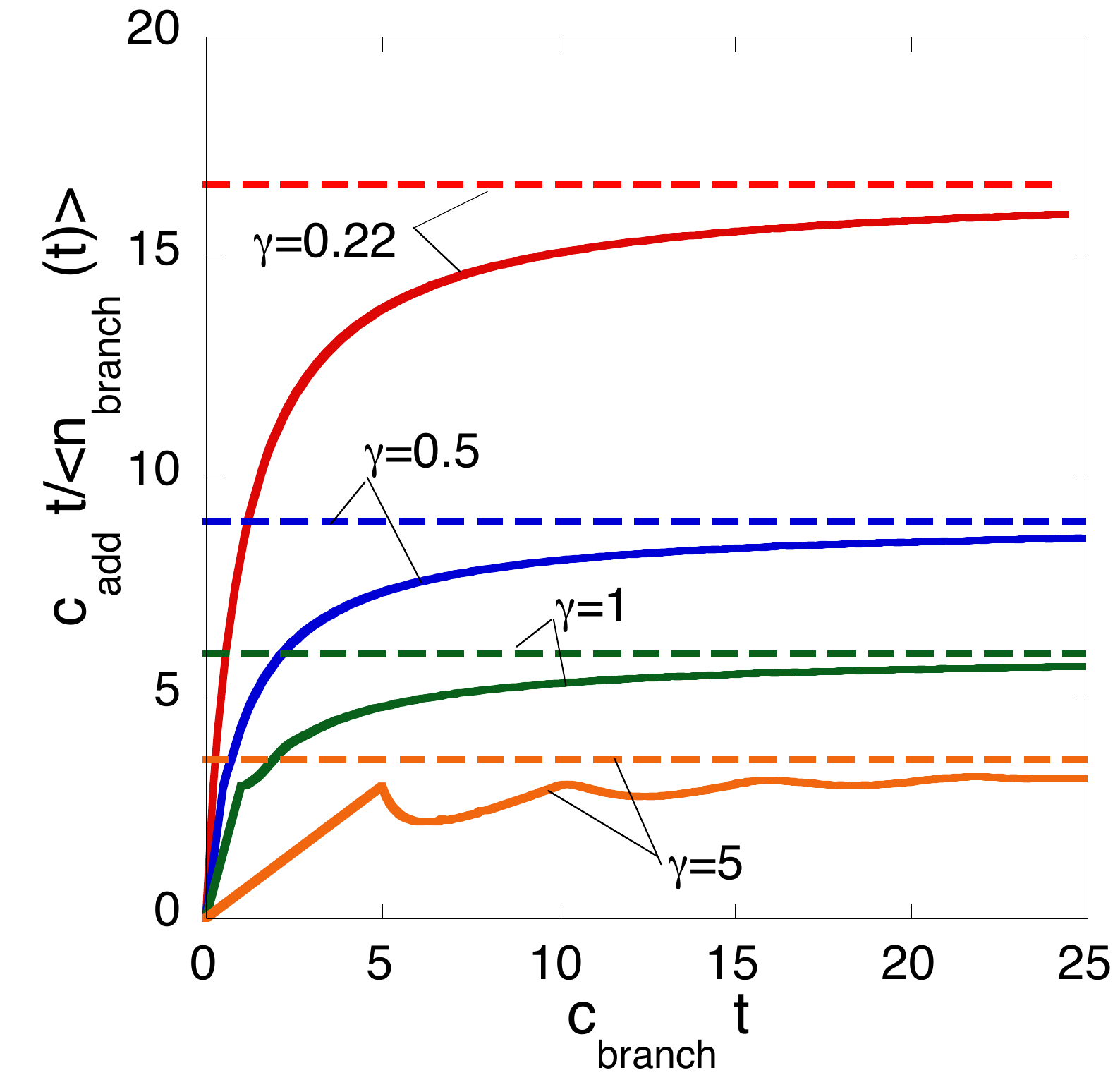}}
\caption{ The mean length of the linear segment, $\la N(t)\ra$, 
  vs. time, for
different values of the parameter  $\gamma=n_0 \frac{c_{branch}}{c_{add}}$ and $n_0=3$.
Also shown by the dashed lines are the large time values, as given by Eq.(\ref{e6}).}
\label{fig:4}
\end{figure}
\begin{figure}
	\centering
		\includegraphics[width=12cm,height=12cm]{{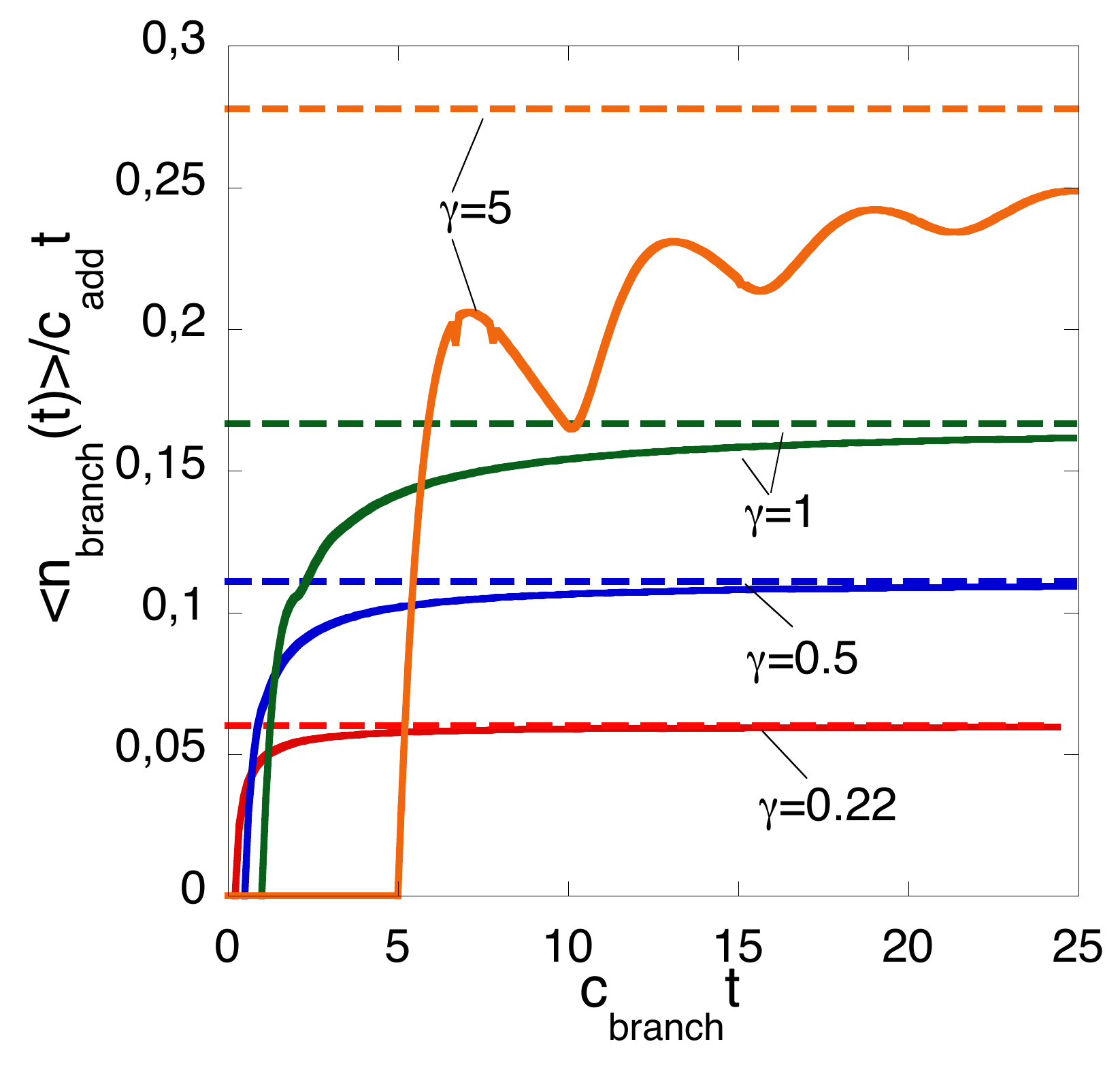}}
\caption{(Color online) The ratio of the mean number of branching to the mean total length of the chain, 
$\la n_{branch}(t)\ra/c_{add}t$ 
  vs. time, for
different values of the parameter  $\gamma=n_0 \frac{c_{branch}}{c_{add}}$ and $n_0=3$.
Also shown by the dashed lines are the large time values as follow from Eq.(\ref{e6}).}
\label{fig:4}
\end{figure}
We note that for $\gamma<<1$ the curved part of the graph in Fig. 6 is due to the fact that branching is delayed relative to growth by about $n_0/c_{add}$, after which it proceeds at a constant rate of 
$\la n_{branch}(t)\ra \approx \tilde{c}_{branch}(t-n_0/c_{add})$. The total length of the chain grows as
$c_{add}t$. For the ratio we, therefore,  have 
 \begin{eqnarray}\label{e14}
\la n_{branch}(t)\ra/c_{add}t\approx \frac{\tilde{c}_{branch}}{c_{add}}-n_0\frac{\tilde{c}_{branch}}{c_{add}^2t}.
\end{eqnarray}
The ratio is  $20\%$ below its large-time value $\tilde{c}_{branch}/{c_{add}}$ at $t\approx 5n_0/c_{add}$, as can be seen from Fig.7. This approximates the range of times in which the initial delay in building a branchable chain leads to non-constant behaviour of the ratio $\frac{number\q of \q branches}{chain \q length}$.
\begin{figure}
	\centering
		\includegraphics[width=12cm,height=6cm]{{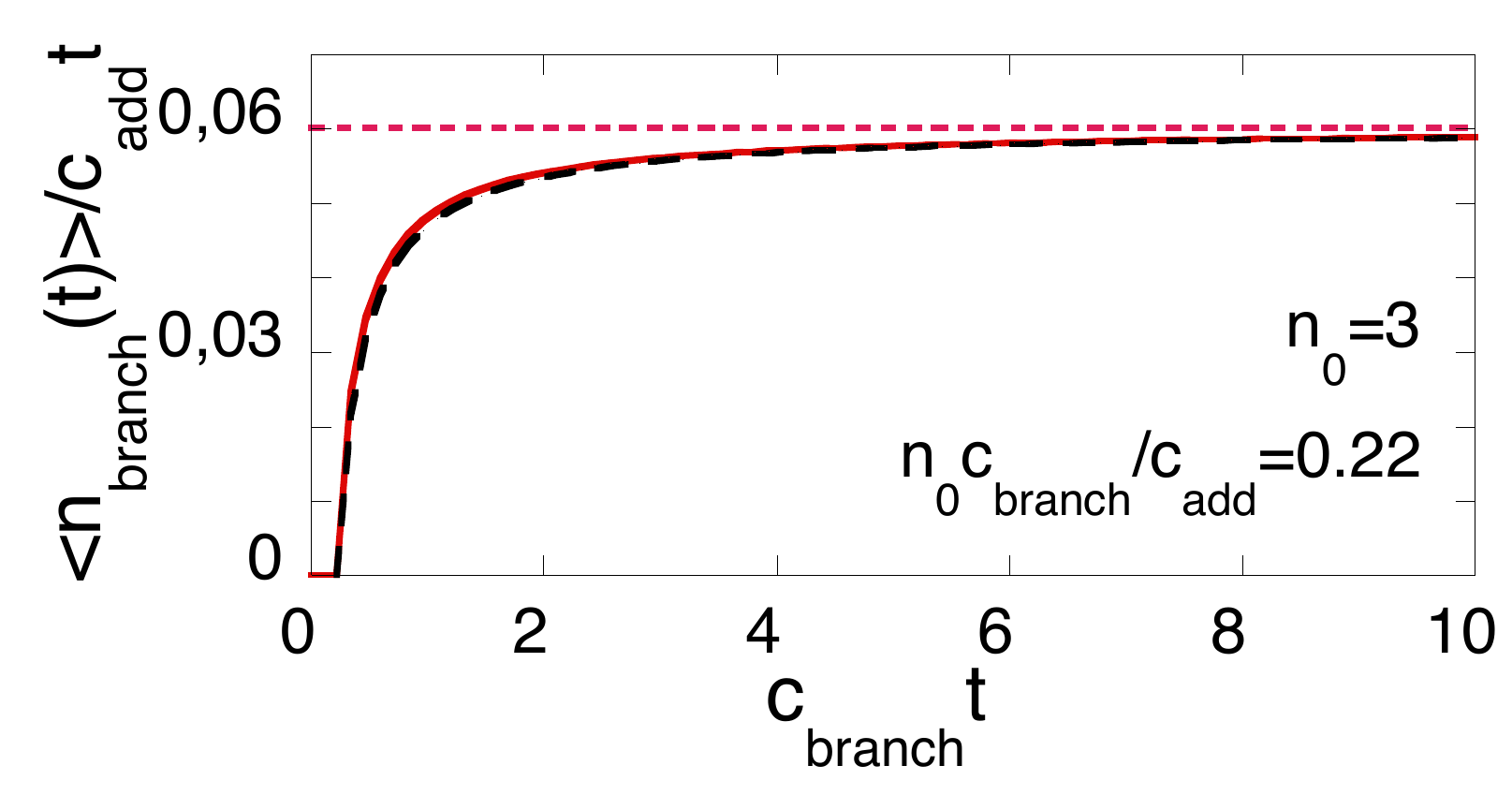}}
\caption{(Color online) The ratio of the mean number of branches to the mean total length of the chain, 
$\la n_{branch}(t)\ra/c_{add}t$, for $\gamma=0.22$: exact (solid) and as given by Eq.(\ref{e14}) (dashed).
Also shown is the large time limit of the ratio (short dashed).}
\label{fig:4}
\end{figure}

\section{Stochastic simulation of delayed growth}
If the growth of the chain does not occur at regular times, but is itself a Poisson process, 
we need to take into account the small probability that three monomers can be added also within a short time. For growth occurring at a constant rate $c_{add}$ the probability to add $n_0$ monomers within a time $\tau$ is 
 \begin{eqnarray}\label{f1}
w_{n_0}(\tau)=
 \int_{0}^\tau d\tau_{n_0-1}...
 \int_{0}^{\tau_2}
 d\tau_1 
 f_{add}(\tau-\tau_{n_0-1})...f_{add}(\tau_1)
= c_{add}^n\frac{\tau^{n_0-1}}{(n_0-1)!}\exp(-c_{add}\tau)
\end{eqnarray} 
where we have used $f_{add}(t)=c_{add}\exp[-c_{add}t]$.
Next we average the delayed PDF for branching, 
$$f_{branch}(t,\tau)\equiv c_{branch}\theta(t-\tau)\exp[-c_{branch}(t-\tau)],$$
over all possible delays, thus obtaining
 \begin{eqnarray}\label{f2}
\bar{f}_{branch}(t,n_0)\equiv \int_0^\infty w_{n_0}(\tau)f_{branch}(t,\tau) d\tau=\n
\frac{c_{branch}c_{add}^{n_0}}{(n_0-1)!}\exp(-c_{branch}t)\frac{d^{n_0-1}}{d\beta^{n_0-1}}\left [\frac{exp(\beta t)-1}{\beta}\right ]_{\beta=c_{branch}-c_{add}}.
\end{eqnarray} 
The shapes of $\bar{f}_{branch}(t,n_0)$ for various values of $n_0$ are shown in Fig.8.
Then we replace $f$ in Eqs.(\ref{a5}) with $\bar{f}_{branch}$, and use them to generate the statistics for branching events. Since we no longer have a simple analytic solution for the resulting non-Markovian equations, we employ a numerical stochastic algorithm similar to that developed by Gillespie \cite{GILL}. 
\begin{figure}
	\centering
		\includegraphics[width=12cm,height=12cm]{{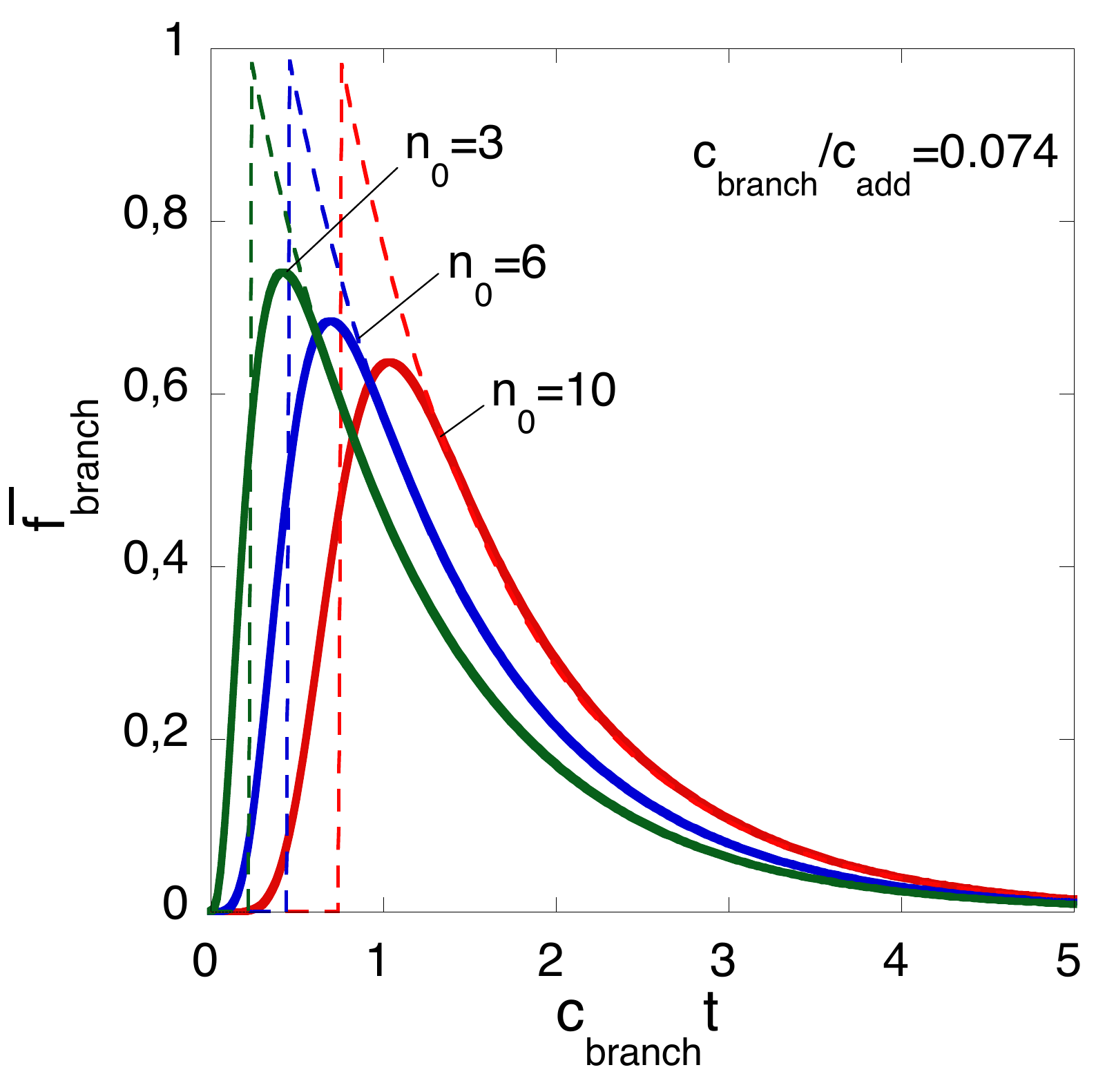}}
\caption{(Color online) The PDF  $\bar{f}_{branch}$ for $c_{branch}/c_{add}=0.074$ for various valued of $n_0$.
Also shown by the dashed lines are the PDFs (\ref{c3}).}
\label{fig:4}
\end{figure}
There are two ways to simulate delayed branching of a polymer chain using the method similar to that of \cite{GILL} and \cite{NMBARC}. A rigorous analysis of their equivalence will be given elsewhere \cite{FUTUR1}.

\subsection { Stochastic simulation of a single delayed process, with the delay built into the corresponding PDF}
A random number generator is prepared, so that it 'draws'  a random number $t_j$, $j=1,2,...K$  with a probability 
 \begin{eqnarray}\label{g1}
\omega_j=f(t_j)dt,
\end{eqnarray}
where $f(t)$ is the PDF given by Eq. (\ref{f2}).
The number of branches is set to zero, and then  the first value  $t_1$ is drawn. If it lies 
between $0$ and the time $t$ at which the growth is stopped, the number of branches is increased by $1$, and $t_1$ becomes the new starting time. This step is repeated until the $k$-th step yields  $t_k>t$, at which point the drawing stops, and the vector $\hat{t} =(t_1,t_2, ..., t_{k-1})$ corresponding to this particular realisation of the system's history is stored. Repeating the simulation a large number of times $N$, one obtains a collection of histories, from which the probability of any particular property can be obtained as the relative frequency with which the property occurs. For example, the probability to have $n$ branches by a time $t$ is given by
 \begin{eqnarray}\label{g2}
P(n,t)=N_n/N, 
\end{eqnarray}
where $N_n$ the number of realisations with exactly $n$ events. 
\newline
The results of a stochastic simulation with the PDF (\ref{c3}), shown in Fig.3, are in full agreement with the analytical results  (\ref{c4})-(\ref{c5}). The results for the PDF (\ref{f2}) are presented in Fig.9.
\subsection { Stochastic simulation of two Poisson processes with an additional constraint}
Alternatively, one can perform a simulation of two simultaneous Poisson processes with their respective PDFs defined as follows
 \begin{eqnarray}\label{g2}
f_{add}(t)=c_{add}\exp(-c_{add}t), \q f_{branch}(t)=c_{branch}\exp(-c_{branch}t), 
\end{eqnarray}
and impose an additional constraint that a branching can only occur after $n_0$ monomers have been added previously. (Note that without such a constraint the processes are independent, and the ratio of the mean number of branchings to the mean length of a polymer in Fig.6 is a constant equal to $c_{branch}/c_{add}$ at all times.)
\newline
Now in each step of the simulation one draws random values of $t_k^{add}$ and $t_k^{branch}$ from the probability distributions $f_{add}$ and $f_{branch}$ in Eqs. (\ref{g2}), respectively.
If $t_k^{add} < t_k^{branch}$, a growth event is recorded at $t_k=t_{k-1}+t_k^{add}$. Otherwise the recorded event is the branching of the chain appearing at $t_k=t_{k-1}+t_k^{branch}$. The step is repeated until 
$t_k$
is found to be greater than $t$, and a history consisting of branching events interspersed among acts of growth is stored. With many histories collected, average values of observables are evaluated as in the previous Subsection. 
\newline
A comparison demonstrates a good agreement between the single- and two-processes simulations of branching events.
The results are shown in Fig.9 for various values of $\gamma$. 
For growth events distributed in time, the sharp features present in the time dependence of observables in the model of Sect. IV, are smoothed over, yet there remain significant non-Markovian effects.
\begin{figure}
	\centering
		\includegraphics[width=12cm,height=12cm]{{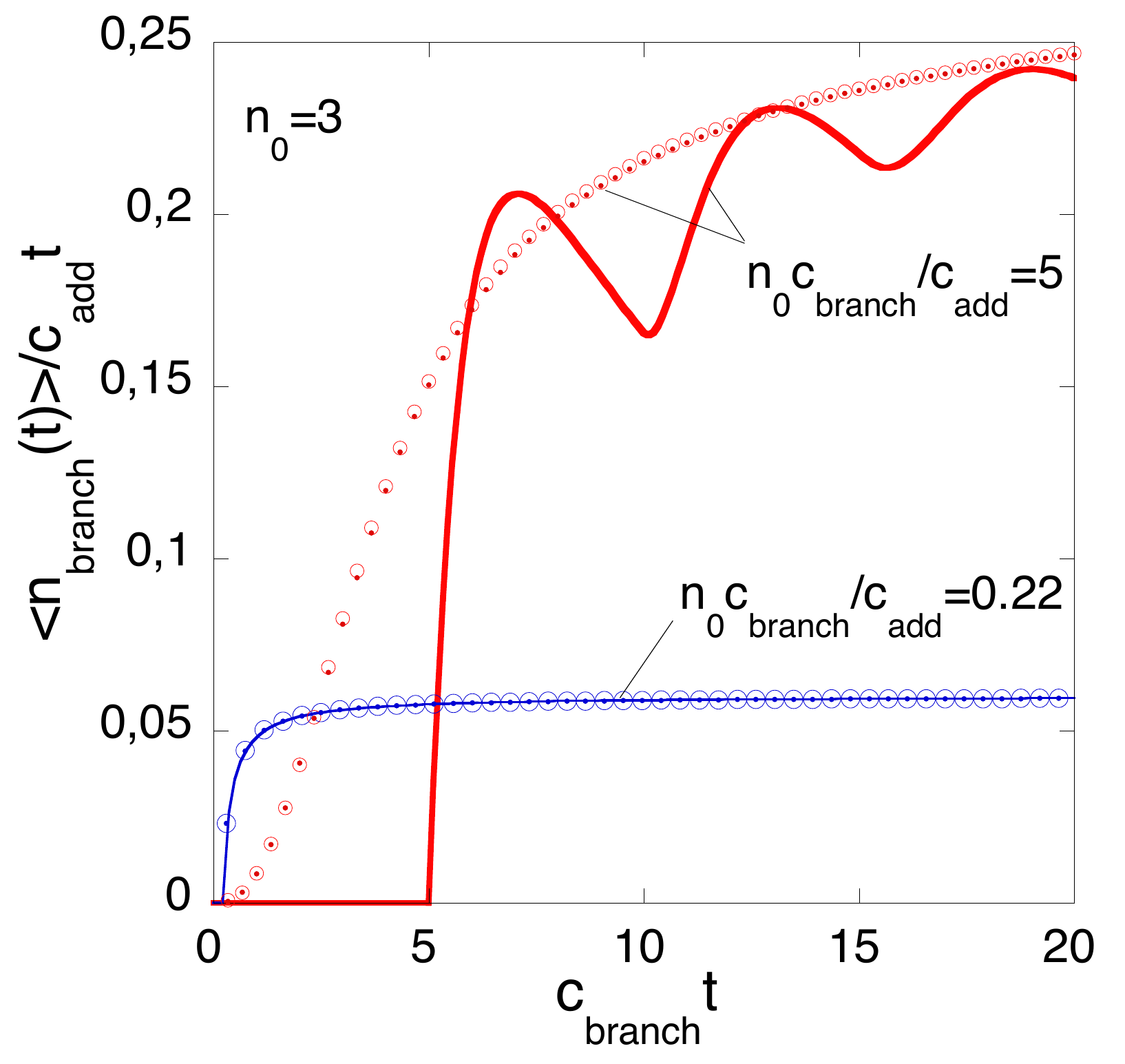}}
\caption{(Color online) The ratio of the mean number of branches to the mean total length of the chain, 
$\la n_{branch}(t)\ra/c_{add}t$ 
  vs. time, for $n_0=3$ and different values of $ c_{branch}/c_{add}$.
  The results for a smooth PDF of Eq.(\ref{f2}) are obtained by the stochastic simulations described in the Subsecs. A (closed circles) and B (large open circles).  The results for PDFs with a uniquely defined delay (\ref{c3}) are shown by solid lines for comparison.}
\label{fig:4}
\end{figure}
\section{Conclusions and discussion}
In summary, introduction of a delay after each growth event significantly changes the statistics of polymer chain growth. The process is governed by the ratio $\tau/t_{add}$, where $\tau$ is the length of the delay, and $t_{add}$ is the time it takes, on average, to add a monomer to the chain. For  $\tau/t_{add}<<1$ the growth remain essentially Markovian, with the mean chain length growing as $t/t_{add}$, except at very short times. For $\tau/t_{add}>>1$ the growth at short times proceeds by rapid attachments of single monomers, separated by long 'waiting periods' in which nothing happens.
As time progresses, the step-like variations of the mean polymer chain become less pronounced, and a growth with a renormalised constant growth rate $\tilde c$ [cf. Eq.(\ref{d4})] is achieved.
\newline
In the simple model of Sect. VI, formation of branches in the chain is a delayed process, whose delay is determined by the growth rate of the polymer chain. The growth itself is unaltered by the branching rate $c_{branch}$ which, in turn, depends on the structural properties of the chain and not on the rate at which the monomers are added. The behaviour of the mean number of branches is, therefore, similar to that of the mean chain length in a linear growth with delay, described in the preceding paragraph. The ratio of the mean number of branchings to the mean length of the chain, $\la n_{branch}(t)\ra/c_{add}t$, has a constant value $c_{branch}/c_{add}$ in the Markovian case. In the presence of a delay, at short times $t\lesssim 10n_0/c_{add}$ it rises from zero to reach the steady 
value of $c_{branch}/(c_{add}+n_0c_{branch})$ as shown in Fig.6.
 In general, the presence of a delay is best visible for $ c_{branch} t_{add} \gtrsim 1$.
\newline
The growth rate $c_{add}=1/t_{add}$ is proportional to the concentration of monomers, and, is therefore, variable.  Our analysis suggests a way of checking whether there are possible time delays of order $\tau$ in the growth of a particular polymer. One way is to examine the mean number of branches occurring at growth times $t\lesssim 5\tau$, in order to see whether the ratio of mean number of branchings to the mean chain length is flat. Admittedly, its experimental realization
may be difficult. 
Alternatively, one might try adjusting $c_{add}$. If for
 $c_{add}\tau \gtrsim 1$ there are  visible deviations from the Poisson statistics of Sect. III, 
 some kind of a  time delay is the likely reason.
 \newline
 To conclude, apart from time delays, there may be various complex processes accompanying  growth of a polymer.  Yet, delayed nature of the growth is one possible reason for observed deviations from the predictions of the Poisson law. 
 \section{Acknowledgements:}
We acknowledge support of the Basque Government (Grants No. IT-472-10, IT-373-10, and  Etortek Nanoiker IE11-304), and of the Ministry of Science and Innovation of Spain (Grant No. FIS2009-12773-C02-01). The SGI/IZO-SGIker UPV/EHU is acknowledged for providing computational resources.

\end{document}